\documentclass[11pt,a4paper]{article}

\AtBeginDocument{%
  \providecommand\BibTeX{{%
    \normalfont B\kern-0.5em{\scshape i\kern-0.25em b}\kern-0.8em\TeX}}}

\usepackage{nccmath}
\usepackage{graphicx}
\usepackage{authblk}
\usepackage[utf8]{inputenc}
\usepackage[english]{babel}
\usepackage{natbib}
\usepackage{booktabs}
\usepackage{url}
\usepackage{hyperref}

\providecommand{\keywords}[1]
{
  \small	
  \textbf{\textit{Keywords---}} #1
} 

\begin{document}

%% anonymous: authors, link to repo, Acknowledgements

%%
%% The "title" command has an optional parameter,
%% allowing the author to define a "short title" to be used in page headers.

\title{Investigating Vulnerabilities of GPS Trip Data to Trajectory-User Linking Attacks}

%%%%%%%%%%%%%%%% Authors' Info %%%%%%%%%%%%%%%%%
%%
%% The "author" command and its associated commands are used to define
%% the authors and their affiliations.

%\author[]{anonymous authors}
\author[1]{Benedikt Ströbl,}
%\email{stroebl@princeton.edu}
\author[2]{Alexandra Kapp}
%\orcid{}
\affil[1]{Princeton University, Princeton, NJ, USA}
\affil[2]{Technische Universität Berlin, Berlin, Germany}

\date{}

\maketitle

\begin{abstract}
Open human mobility data is considered an essential basis for the profound research and analysis required for the transition to sustainable mobility and sustainable urban planning. Cycling data has especially been the focus of data collection endeavors in recent years. Although privacy risks regarding location data are widely known, practitioners often refrain from advanced privacy mechanisms to prevent utility losses. Removing user identifiers from trips is thereby deemed a major privacy gain, as it supposedly prevents linking single trips to obtain entire movement patterns.
In this paper, we propose a novel attack to reconstruct user identifiers in GPS trip datasets consisting of single trips, unlike previous ones that are dedicated to evaluating trajectory-user linking in the context of check-in data. We evaluate the remaining privacy risk for users in such datasets and our empirical findings from two real-world datasets show that the risk of re-identification is significant even when personal identifiers have been removed, and that truncation as a simple additional privacy mechanism may not be effective in protecting user privacy. Further investigations indicate that users who frequently visit locations that are only visited by a small number of others, tend to be more vulnerable to re-identification. 
\end{abstract}

\keywords {trajectory-user linking, privacy, mobility data, GPS trip data}

\section{Introduction}

Mobility data, including human movement patterns, is considered essential for the just transition to sustainable mobility~\citep{schneider_mobility_2023, nelson_crowdsourced_2021} and an increasing number of crowdsourcing approaches collect such data. For example, the smartphone app \textit{CycleTracks}\footnote{\url{https://www.sfcta.org/tools-data/tools/cycletracks}} by the San Francisco County Transportation Authority enables users to track bicycle trips to help understand the needs of cyclists. Commonly, such datasets are only available for researchers or local authorities,
thereby, excluding certain stakeholders~\citep{oksanen2023geoprivacy}, such as startups or the civil society, and additionally, obstruct data access for intended users. 
Thus, there are increasing efforts to provide such crowdsourcing data \textit{openly available}, for example, the GeoPrivacy platform~\citep{oksanen2023geoprivacy},
the OpenBikeSensor\footnote{\url{https://www.openbikesensor.org/}}, or
SimRa~\citep{karakaya_simra_2020}, a smartphone app collecting near-miss accidents.

However, the provision of open mobility data can pose major privacy risks, as home and workplace can easily be inferred~\citep{yangDetectingHomeWork2021}, and sensitive insights may be gained based on visited locations, like demonstrations or places of worship.
 The high degree of uniqueness of human movement patterns poses an enormous challenge in anonymizing mobility data \citep{demontjoyeUniqueCrowdPrivacy2013b} since little outside information is sufficient for re-identification even when obvious personal identifying information such as name or phone number has been removed \citep{krummInferenceAttacksLocation2007b,culnane_stop_2019, zang_anonymization_2011}. For example, Culnane et al. \citep{culnane_stop_2019} re-identified different individuals in a supposedly anonymized smart card dataset with only little publicly available information as side information, retrieving complete movement patterns spanning multiple years.

Considering a mobility dataset where each record consists of a GPS trajectory like a bicycle trip:
%\footnote{Henceforth, the terms 'trajectory' and 'trip' are used interchangeably.}:
the more trips an individual contributes to the dataset, the \textit{higher their risk for re-identification}, as more information is available to link side information. Simultaneously, given a successful re-identification attack, the \textit{privacy invasion} increases with the number of contributed trips.
Broadly speaking, knowing every location a person has traveled to for an entire year is more critical than only linking a single trip to a person.
Both considerations, the re-identification risk and the extent of the privacy invasion, are based on the assumption that records within a dataset are \textit{linked with one another} based on a (pseudonymized) user identification.
Consequently, a straightforward privacy enhancement is the entire omission of a user-level ID.
In fact, omitting the user ID is practiced in the industry for privacy enhancement ~\citep{kapp_collection_2022-1} where simple techniques that retain higher utility are typically preferred to more advanced privacy concepts such as \textit{k}-anonymity \citep{sweeneyKanonymityModelProtecting2002} and differential privacy \citep{dworkDifferentialPrivacySurvey2008}. Also, open data endeavors like GeoPrivacy~\citep{brauer_human_2023} or SimRa~\citep{karakaya_simra_2020} make use of this method.

In light of rudimentary mechanisms being widely used in practice, it is imperative to gain a more holistic understanding of their efficacy. While there is a plethora of research on re-identification attacks of trajectory micro-data, i.e., trip data, (cf.~\citep{fiorePrivacyTrajectoryMicrodata2020a}), to the best of our knowledge there is no study investigating the privacy provided by omitting the user ID of trip datasets.
Given the highly unique, yet recurrent nature of individual movement patterns the question arises as to whether such user IDs could be reconstructed by an attacker, thereby raising the privacy risk substantially.

Linking trajectories based on the common user is not a new concept.
\citet{gaoIdentifyingHumanMobility2017} introduce the term \textit{trajectory-user linking} (TUL) which aims at linking trajectories back to users in the context of location-based social network (LSBN) data where users shared their locations by checking in. In their experimental setup sub-trajectories are created by segmenting users' check-in sequences according to 6-hour time windows. The subsequent objective is to reconnect these segmented sub-trajectories. 
Since its first introduction, a series of papers have investigated TUL with various semi-supervised deep learning architectures to improve functionalities such as personalized recommendations or crowd monitoring \citep{zhouTrajectoryUserLinkingVariational2018, miaoTrajectoryUserLinkingAttentive2020, sunTrajectoryUserLinkAttention2021, chenMutualDistillationLearning2022}. Different works consider TUL in the context of an attack and propose a model to resist it\citep{lun_resisting_2023, liTulam2023,raoLstmTrajgan2020, korichiLeveraging2024}. However, similar to previous TUL publications they also consider trajectories consisting of chunks of LSBN data. Consequently, the characteristics of these trajectories highly diverge from trips that connect an origin with a destination, especially in terms of time span and the number of semantic staypoints. 
Thus, we see a gap in the literature regarding the re-identification risk in trip datasets where the user ID has been omitted. This publication provides the following contributions:

\begin{itemize}
    \item We formulate the TUL attack in the context of GPS trips.
    %\item We formulate a novel problem for mobility privacy research with respect to reconstructing user identifiers in datasets with no trajectory-user link.

    \item We develop a new attack solving the aforementioned task by leveraging a set of assumptions about the mobility routines of urban residents and assigning trajectories assumed to be from the same person to a common identifier.
    
    \item We evaluate the privacy issue inherent to GPS trip data with experimental studies and find high variability in re-identification risks with substantial implications for a significant fraction of users.
    
    \item We evaluate the effectiveness of trajectory truncation as an obfuscation technique highlighting its unreliability across real-world datasets.
\end{itemize}

This paper is structured as follows: We first introduce our methods in Section \ref{chapter:methods}, encompassing the problem formulation, the proposed attack, and a common obfuscation technique which is evaluated according to its susceptibility against our attack.  We continue in Section \ref{chapter:experiment} with our attack evaluation and discuss our results in Section \ref{chapter:discussion}.

\section{Methods}
\label{chapter:methods}

Our attack aims to reconstruct user IDs in GPS trip datasets with no trajectory-user link by combining a set of assumptions about the day-to-day mobility patterns of urban residents. In accordance with \citet{gaoIdentifyingHumanMobility2017}, we first formally define the underlying task as a clustering problem before we introduce our approach to solving it as an attack.

\subsection{Problem Formulation}

Let $T = \{T_1, \dots, T_m\}$ be a set of $m$ GPS trips generated by a set of $s$ users $U = \{u_1, \dots, u_s\}$ ($m \geq s$) represented by some set of IDs, where each trip $T_{k} = \{l_{1}, l_{2}, \dots, {l_{n}}\}$ is a temporally ordered sequence of $n$ spatio-temporal points $l_{j}$. Each point $l_{j}$ consists of coordinates (i.e., latitude and longitude) recorded at time $t_j$. The trips are \textit{unlinked}, i.e. there is no mapping from $T$ to $U$. The objective is to reconstruct a user ID linking trips produced by the same user. Hence, an optimal solution to this problem is to find a mapping $T \to U$ such that every trip $T_k \in T$ is correctly assigned to the ID of the user it was produced by.

\subsection{Reconstructing User IDs}

In the following, we introduce a method to solve the outlined problem with a series of heuristics. These include combining trips that seem to be continuations of one another and identifying potential home locations of individuals. In addition, we leverage the information content of location co-visits to combine initially identified groups of trajectories and refine the cluster assignments. 

Moreover, since the goal of this method is to assess the feasibility of reconstructing an assignment of trajectories produced by the same person to a common user ID, it is important to point out that the ground-truth user IDs of both datasets are only considered during the evaluation but never throughout the attack itself.

\subsubsection{Trip Concatenation}

\citet{najjarTrajectoryUserLinkingEasier2022a} investigated the high uniqueness of visit patterns among users in various datasets and found that over 99\% of all visited locations in their data were uniquely visited by one person. Additionally, the uniqueness of a person's mobility trace has been studied extensively across mobility datasets of various types \citep{rossiSpatiotemporalTechniquesUser2015, demontjoyeUniqueCrowdPrivacy2013b, xuTrajectoryRecoveryAsh2017a}. Thus, we hypothesize that if two trips arrive and begin in the direct vicinity of one another with the departing trip being a unique candidate starting within $h_{\mathrm{concat}}$ hours after the arriving trip ended, it is likely that these are part of a continued trip of the same user. Therefore, these two should be assigned to the same user ID.

To detect trip continuations, we map the start and end points (SP and EP) of trips onto a grid where each rectangular grid cell has a side length $s_{cell}$. We thus expect a similar cell ID for the end of the first and the start of the second trip of a continued trajectory.
To prevent mismatches, two trips are only linked if there are no other trips arriving $h_{concat, before}$ hours before and $h_{concat, after}$ hours after the arrival time of a candidate trip at the same cell.

\subsubsection{Home Location Assignment}

The home of a person is one of the most frequently and routinely visited places \citep{hasanSpatiotemporalPatternsUrban2013} and thus an effective starting point for re-identification in mobility datasets~\citep{golle2009anonymity,freudiger2012evaluating, krummInferenceAttacksLocation2007b}.
Therefore, we set out to find a set of potential \textit{home locations} (HL) of the users as they are expected to be highly unique to and highly frequented by the same person. Trajectories starting or ending in the same HL could consequently be clustered to the same preliminary user ID.
From research investigating energy demand \citep{torritiUnderstandingTimingEnergy2017} and commuting patterns in cities \citep{maUnderstandingCommutingPatterns2017, choFriendshipMobilityUser2011b}, it has become evident that most urban residents leave home between 6 and 10 AM and only return after 6 PM but before midnight. Hence, we define two disjoint time intervals $t_{morning}$ and $t_{evening}$ for which all grid cells where a trip in our data started during $t_{morning}$ or ended during $t_{evening}$ are defined as potential HLs. Similar to the concatenation step described in the previous section, we exclude cells from defining an HL where more than one trip starts during $[h_{morning, before}, h_{morning, after}]$ or ends during $[h_{evening, before},$ $h_{evening, after}]$, respectively. We note that this implicitly introduces a simplification such that a correct assignment for more than one user living in the same home location is prevented.
As illustrated in Appendix \ref{Figure:dissolvingexample}, we dissolve adjacent HL cells resulting from the previous step of defining HL candidates to one combined polygon to account for the fact that -- if they are to come from the same person -- these are likely due to measurement variance inherent to GPS \citep{louMapmatchingLowsamplingrateGPS2009} and to keep the overall number of HLs small\footnote{Note here that the cell side length $s_{cell}$ is set to 200m}. 

\begin{figure}[h]
    \centering
    \includegraphics[width=\linewidth]{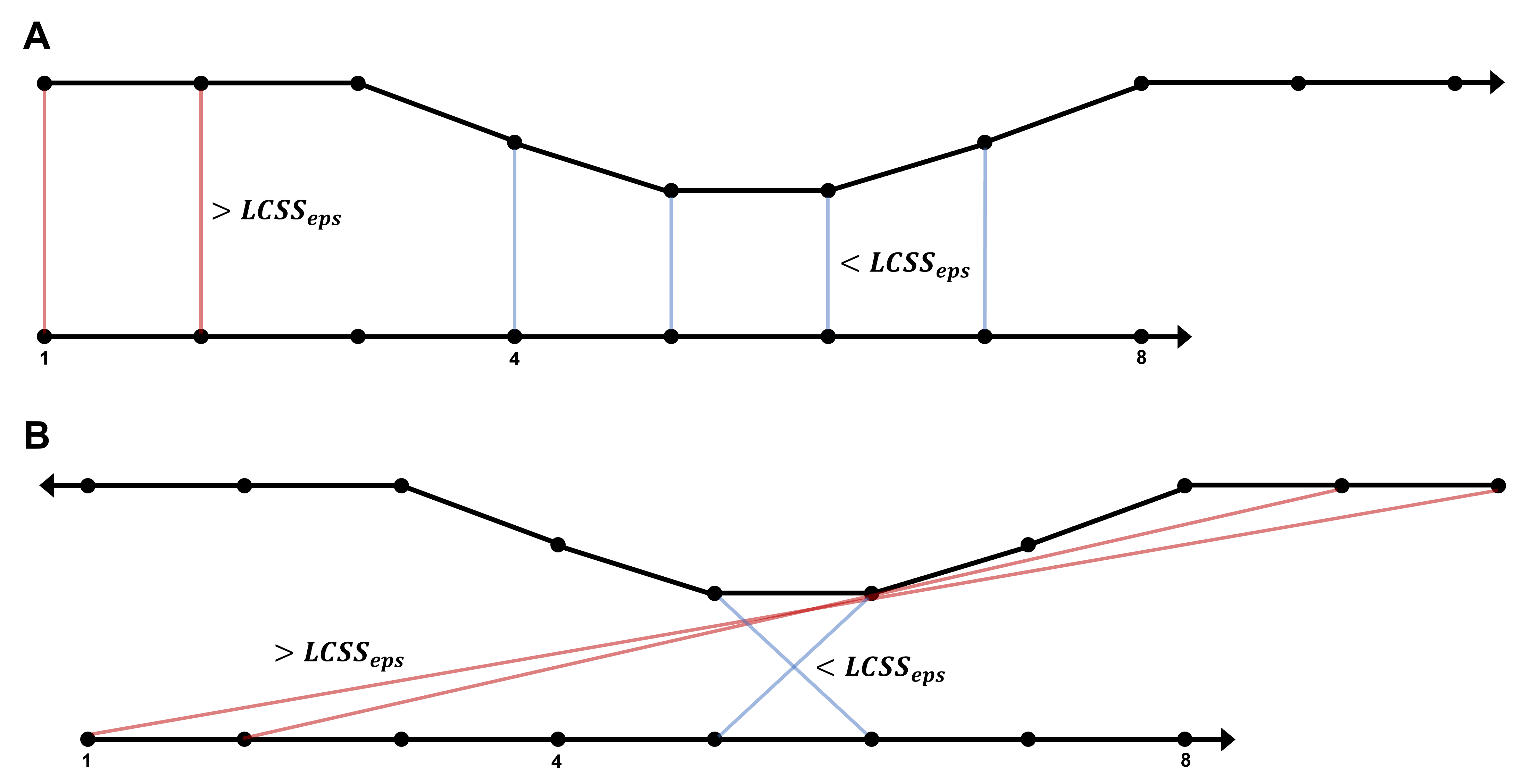}
    \caption{Example illustration of LCSS metric for a pair of trips. (A) being the original trip and (B) with the upper trip reverted. The LCSS value in (A) is 0.5 since 50\% of the points of the shorter trip are within a distance threshold $LCSS_{\varepsilon}$ from their counterpart in the longer trip. Analogously, in (B) the LCSS results in 0.25 because when comparing the points in reverted order -- the direction of travel for one of the trips goes in the opposite direction, only 25\% of the points are within $LCSS_{\varepsilon}$. Note here that for easier readability the red lines indicating points that are further apart than the specified threshold, are only plotted exemplary for two of the above point pairs.}
    \label{Figure:lcss}
\end{figure}

\begin{figure*}[h]
    \centering
    \includegraphics[width=\textwidth]{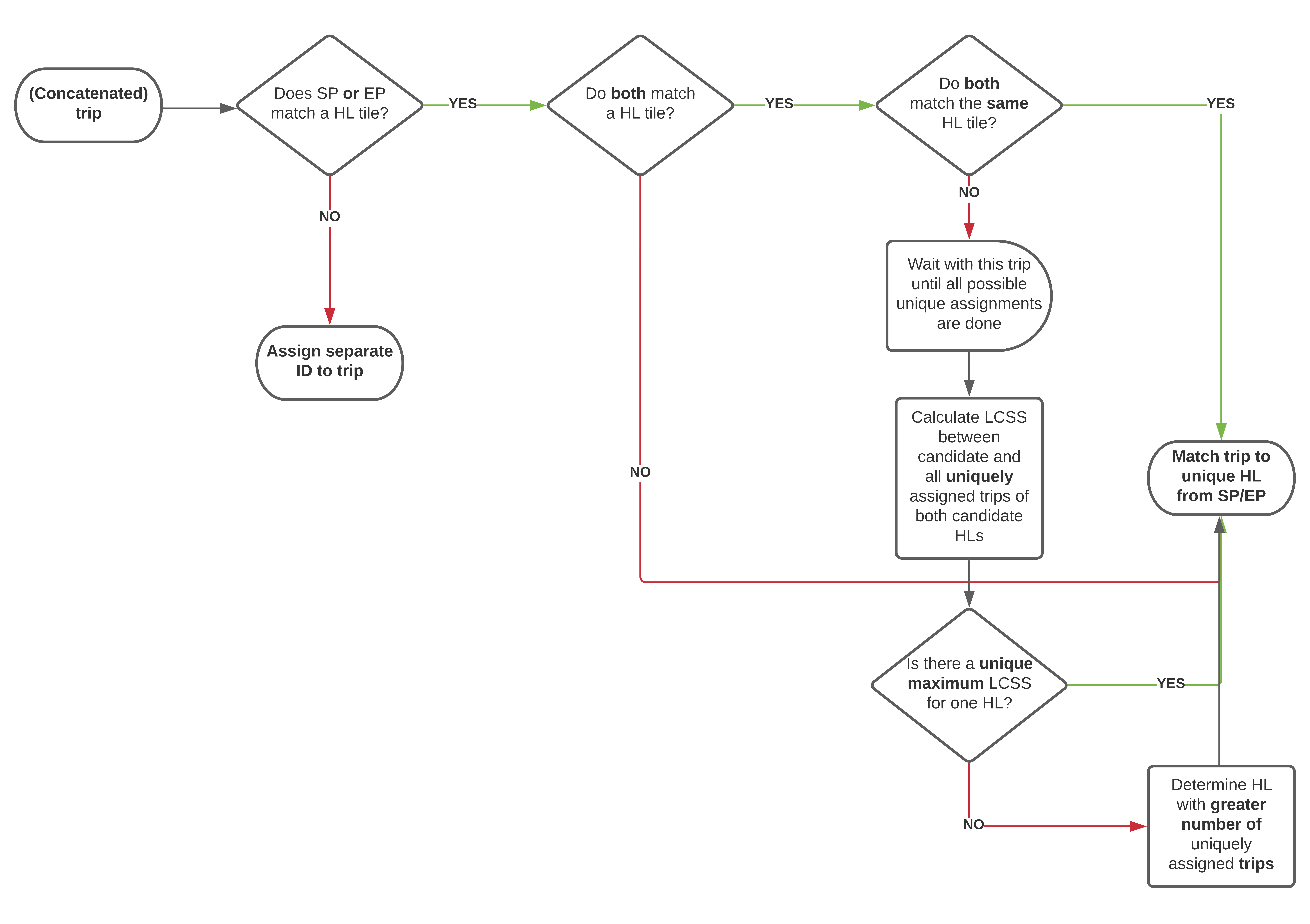}
    \caption{Flowchart explaining home location assignment procedure for concatenated trips}
    \label{Figure:flowchart_hl}
\end{figure*}

After defining HL candidates, trips are clustered based on common HLs. However, a trip can start and end in two different HL candidates. To resolve such double matches, we apply the procedure shown in Figure \ref{Figure:flowchart_hl}, which is as follows. When a trip begins and ends in two distinct HL cells, we compute the similarity of said trip with all uniquely assigned trips of both HLs using their Longest Common Subsequence (LCSS) \citep{vlachosDiscoveringSimilarMultidimensional2002}. LCSS values range from 0 to 1, where 1 would mean that two trips fully match. For example, an LCSS of 0.5 indicates that 50\% of the points part of the shorter trip can be considered equivalent -- their distance is less than $LCSS_{\varepsilon}$ -- to their counterparts in the longer one while traversing monotonically from start to end (see Figure \ref{Figure:lcss}). Since we are agnostic about the direction of travel when comparing the similarity of two trips, we also apply LCSS after reverting the direction of one of the trips and take the maximum as the final result. This metric has been shown to be robust against not too widely varying sampling rates and sequence lengths \citep{tooheyTrajectorySimilarityMeasures2015} and allows matching these trips to the HL candidate with the greatest unique LCSS value.

If the comparison of the two highest LCSS values of the candidate HLs does not lead to a definite maximum, we keep comparing the next highest LCSS values until one is found. In case there still exists no distinct maximum after comparing all the LCSS values of the two HLs, we assign the trip to the HL with the greatest number of uniquely assigned trips. Trips that have no overlap with any of the HLs, are each assigned a separate ID. All formerly concatenated trips are assigned to the same HL.

Lastly, after assigning each trip to an HL, we take into account that no person can make two trips at the same time, and compute the largest common subset of non-simultaneous trips assigned to the same HL and assign a separate preliminary user ID to each trip not being part of this set.

\subsubsection{TF-IDF Refinement}

By concatenating trips likely to be continuations of one another and assigning these to potential HLs we have already assigned a preliminary user ID to each trip. To further refine these preliminary clusters according to a common user ID, we consider the location visit patterns in a subsequent step. 

An extensive amount of research has been done on the high degree of uniqueness and strong predictive power of a person's location visits in mobility datasets \citep{demontjoyeUniqueCrowdPrivacy2013b, najjarTrajectoryUserLinkingEasier2022a, pappalardoAnalyticalFrameworkNowcast2016}. In that line of research, \citet{gonzalezUnderstandingIndividualHuman2008} have put forward that human mobility follows clear patterns that are often easy to reproduce indicating that people tend to routinely visit the same places.

Building on this, an intuitive approach to capturing the information content of correlating visitation patterns between two sets of trips is Term Frequency Inverse Document Frequency (TF-IDF). TF-IDF is a widely-used statistic in text analysis \citep{rajaramanDataMining2011} that measures the relative importance of terms in documents part of a larger collection of texts. In the realm of mobility data, adapted versions of TF-IDF have for instance been applied to measure the relevance of road segments \citep{mahrsiModularityBasedClusteringNetworkConstrained2012a} or extracting urban zone embeddings \citep{yaoRepresentingUrbanFunctions2018}. For the present case, we use an adapted version of this metric to quantify the relative rarity of commonly visited locations between two preliminary trip clusters. Consequently, we use TF-IDF to measure the importance of different locations (`term') to a preliminary group of trips (`document') that we assume to be from the same person. Hence, for the remainder of this definition, we will refer to the preliminary clusters of trips having been assigned the same ID as belonging to a single `user'. 

Different from the previous steps, we discretize the space within the defined spatial bounding box with a \textit{new} set of grid cells $G$ in which each rectangular cell $g \in G$ now has a different side length $s_{cell, tfidf}$ that is set to 500m in order to get a smoother distribution of visits across all cells. Mapping all trips' SPs and EPs of a user $u$ onto $G$ creates a set of visited cells $G_u$ and their corresponding visit frequencies. Accordingly, the resulting TF-IDF statistic is defined as in Equation \ref{equation:tfidf} with $tf(g, u)$ being the relative frequency of the raw count $f_{g, u}$ with which user $u$ visited location $g$ in relation to the total number of location visits in the denominator (see \ref{equation:tf}). $idf(g, U)$ defined in Equation \ref{equation:idf} denotes the log-scaled inverse fraction of users that visited location $g$.

% \vspace{20pt}
\begin{ceqn}
\begin{align}\label{equation:tf}
    tf(g, u) = \frac{f_{g, u}}{\sum_{g' \in G_u} f_{g', u}}
\end{align}
\end{ceqn}

%\vspace{12pt}
\begin{ceqn}
\begin{align}\label{equation:idf}
    idf(g, U) = log\left(\frac{|U|}{1 + |\{u \in U:g\in G_u \}|}\right)
\end{align}
\end{ceqn}

%\vspace{12pt}
\begin{ceqn}
\begin{align}\label{equation:tfidf}
    tfidf(g, u, U) = tf(g, u) \cdot idf(g, U)
\end{align}
\end{ceqn}
\vspace{3pt}

With $tfidf(g, u, U)$ calculated for all users and across all location visits, we construct a matrix populated with the location similarity metric as defined in Equation \ref{equation:locsim} for every possible pair of users $u_i$ and $u_j$. The overall location similarity between $u_i$ and $u_j$ is the mean interaction of TF-IDF statistics across all co-visited locations of the two users denoted as $G_{u_i, u_j}$.

%\vspace{20pt}
\begin{ceqn}
\begin{align}\label{equation:locsim}
    locsim(u_i, u_j) = \frac{1}{|G_{u_i, u_j}|}\sum_{g\in G_{u_i, u_j}} tfidf(g, u_i, U) \cdot tfidf(g, u_j, U)
\end{align}
\end{ceqn}
%\vspace{3pt}

Based on the location similarity matrix, we iteratively realize a number of $n_{matches}$ best matches -- effectively combining two user IDs into one -- before recalculating the similarity matrix. This process is repeated until a predefined stopping criterion is reached. Assuming that an attacker would only want to realize those matches that have a particularly high location similarity in order to minimize mistakes, we define this stopping criterion as a fixed threshold relative to the TF-IDF values across all users and locations calculated during the first iteration. In other words, the procedure is stopped if the best possible match of an iteration has a location similarity measure that is less than the squared $q_{match}$-percent quantile of all TF-IDF scores calculated during the first iteration. For example, setting $q_{match}$ to 0.5 would mean that we only continue to combine user IDs if the location similarity defined in Equation \ref{equation:locsim} between two users is greater than what we would expect for a co-visited location that had only median TF-IDF during the first iteration. 
It is important to point out that we are not applying standard clustering techniques to the TF-IDF weighted location visit patterns to further refine our user ID mapping since the objective is not to find overall similar sets of trips but rather the highly relevant link that contains -- from an information-theoretic view -- high `amounts of information' \citep{aizawaInformationtheoreticPerspectiveTf2003}. In other words, this approach aims to link two otherwise disjoint sets of trips that reflect two behavior patterns of the same person. For example, instead of matching geospatially similar trips, such as all the commuters traveling to a business district starting in the same part of town, we identify if there is a trip connecting the HLs of an individual in our data and their hypothetical partner in case this person frequently stays over at their partner's apartment.

\subsection{Obfuscation Technique}

Openly available GPS mobility datasets often include more privacy measures than only the omission of a user ID. Thus, we will evaluate the success of our proposed attack given such a typical countermeasure.
Location-based applications, such as fitness tracking apps \citep{dhondtRunDayWon2022a, hassanAnalysisPrivacyProtections2018a} and SimRa, commonly provide functionality of privacy zones or trajectory truncation which both aim at removing data from sensitive locations, typically at the origin and destination of a trip \citep{EditMapVisibility2023, ConnectUnderstandingPrivacyZones, SimRaSicherheitIm}. In this study, we evaluate the efficacy of truncating trips as an obfuscation technique and focus on the implementation of this measure used by the SimRa project. 

\begin{figure}[h]
    \centering
    \includegraphics[width=5cm]{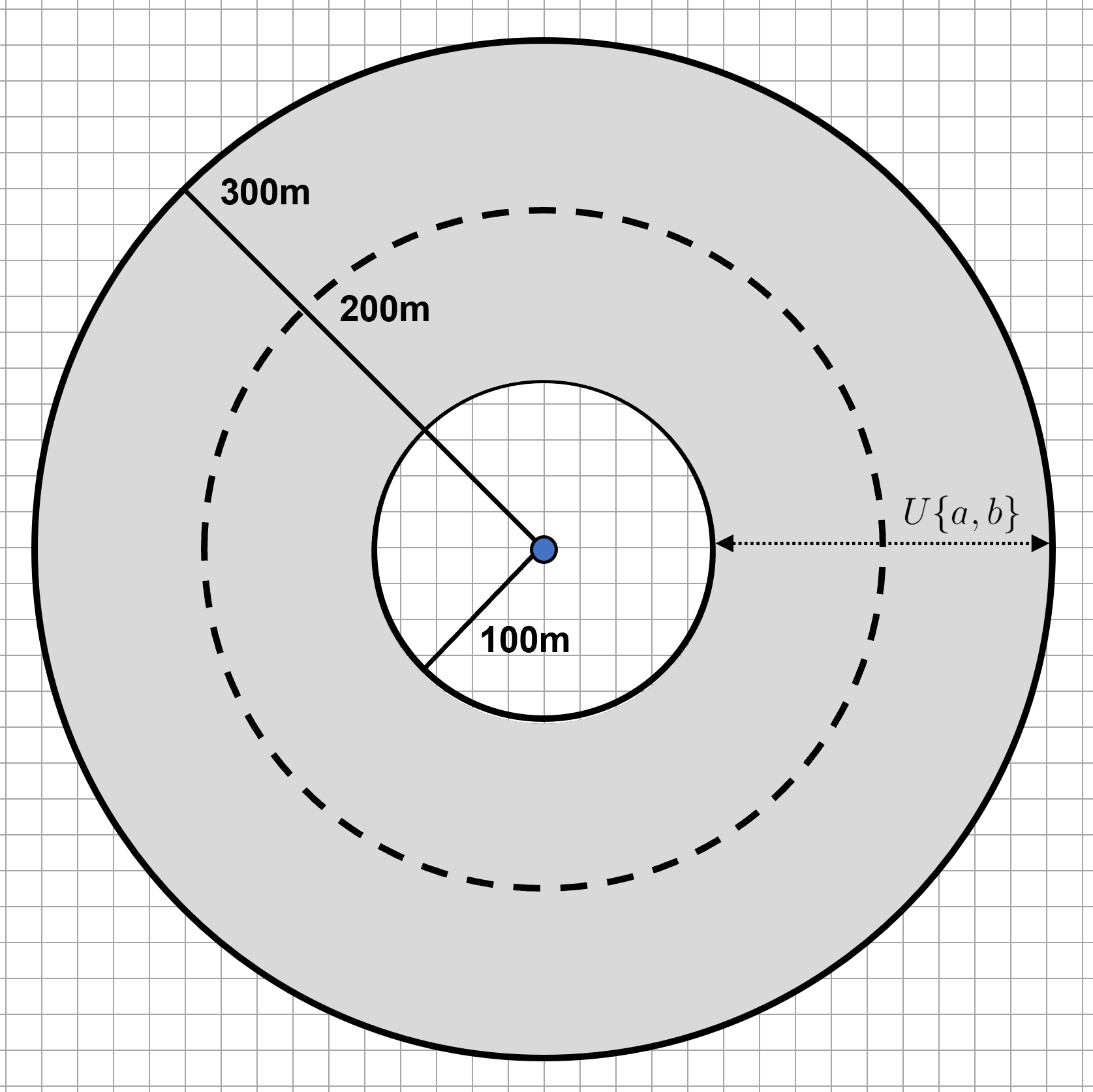}
    \caption{Obfuscation technique applied to conceal sensitive SPs and EPs of trips. Radius for truncation is drawn from uniform distribution $U\{a,b\}$ with $a = 100m$ and $b = 300m$.}
    \label{Figure:obfuscationtechnique}
\end{figure}

SimRa users have the option to truncate their trajectories by a variable margin set with a slider control around SP and EP. However, the default is to upload the original GPS recording without any obfuscation applied. In the app's preferences, users can modify this setting in order to always hide an initial segment of up to $50m$ by default (refer to Appendix \ref{appendix:simra} for screenshots of the SimRa user interface). For our evaluation, we assume a privacy-sensitive community that truncates the ends of \textit{every} trip by a mean radius of $200m$, which is in agreement with the default privacy setting on fitness tracking apps\citep{dhondtRunDayWon2022a}. To account for the fact that users can choose an arbitrary truncation margin in the SimRa app, we sample radii from a uniform distribution between $100m$ and $300m$ for each trip as illustrated in Figure \ref{Figure:obfuscationtechnique}.

\section{Experimental Evaluation}\label{chapter:experiment}

\subsection{Dataset Description}
\label{chapter:datasets}

We evaluate the attack based on two datasets, namely the widely-used \textit{GeoLife} \citep{zheng2011geolife} and a new non-public dataset collected by the \textit{DLR MovingLab} in the context of the \textit{freemove}\footnote{\href{https://www.freemove.space/en/}{https://www.freemove.space/en/}} research project, hereafter referred to as \textit{freemove}. While these datasets contain other variables, this study only includes user ID, latitude, longitude, and timestamp information, with all additional information being discarded
\footnote{The accompanying GitHub repository can be accessed via 
\url{https://github.com/benediktstroebl/Master-Thesis}}.
%\url{https://anonymous.4open.science/r/TUL-attack-CB53}}.

\textit{freemove} is a dataset containing 1,408 trajectories donated by 74 participants of a research project who were mainly students recruited through public announcements and on campuses of Berlin universities. The study was conducted over two weeks from 10.31.2022 to 11.13.2022. 

The \textit{GeoLife} \citep{zheng2011geolife} dataset collected by Microsoft Research Asia contains 17,761 GPS trajectories of 182 users contributed over a total of five years between April 2007 and August 2012. The data was recorded by different GPS loggers and phones capturing a diverse range of users' movements in day-to-day life, such as commuting, shopping, dining, and sports activities. Following the approach of \citet{rossiSpatiotemporalTechniquesUser2015, luTrajectorySplicing2020}, we limit the data to records tracked in 2008 since this is among the most active periods in the data.

Both datasets are not uniformly distributed across their cities but show a large spatial overlap of many participants; likely due to the high share of TU students in the freemove and Microsoft employees in the GeoLife dataset. It is important to stress that this does \textit{not} reduce the complexity of re-identifying individuals because this \textit{increases} the spatio-temporal overlap between users' movements.

Both datasets are preprocessed as follows, to account for measurement errors and artifacts in the data. Trajectories that are shorter than 200 meters or have less than 50 GPS recordings are discarded. This removes data points that are the result of users accidentally starting to record or GPS connection issues that produced trips with only a few GPS points. Further, we remove the longest 5\% of all trips in both datasets to account for extreme outliers that are likely the result of users not ending the recording after a trip has been completed. 
The analysis is restricted to trips within the urban areas of Berlin (bounding box: 52.100, 12.562, 52.803, 14.129) and Beijing (bounding box: 39.600, 116.080, 40.270, 116.690) each including a generous buffer around the suburbs to mainly filter long-distance trips, e.g., flights, hence excluding only trajectories outside of the main city areas. This yields a total of 1,186 trips from 74 users and 5,101 trips from 73 users for freemove and GeoLife, respectively. In Table \ref{tab:datasummary}, a basic overview of the two resulting datasets is provided using a five-number summary.

Finally, Table \ref{tab:parameters} shows the parameter values used in this study during evaluation for all variables introduced throughout Section~\ref{chapter:methods}.

\subsection{Evaluation Metrics}

To quantify the \textit{overall} performance of the applied attack across users, we draw on standard clustering evaluation metrics and compare the reconstructed user IDs with the ground truth. Considering that the number of identified clusters can get very large compared to the true number of users in our datasets, we report the Adjusted Rand Index (ARI) \citep{hubertComparingPartitions1985, steinleyPropertiesHubertArableAdjusted2004} and the Adjusted Mutual Information (AMI) \citep{vinhInformationTheoreticMeasures2009} that have both been conceived to adjust for random labeling \citep{vinhInformationTheoreticMeasures2009}. ARI is bounded between -0.5 for especially discordant clusterings and 1 when they are identical to the ground truth. AMI, on the other hand, is upper limited by 1.0 if the clustering and ground truth are identical and it has an expected value of 0 in case of a random partitioning. In addition, we report the homogeneity and completeness scores as defined in \citet{rosenbergVMeasureConditionalEntropyBased2007a} since they serve as an intuitive information-theoretic interpretation of the `type' of mistakes made by the outlined approach. These metrics each range from 0 to 1, where a completeness score of 1 signifies that all trips of an individual are assigned to the same cluster, and a homogeneity score of 1 means that each cluster contains only trips from the same person.

\begin{figure}[h]
    \centering
    \includegraphics[width=\linewidth]{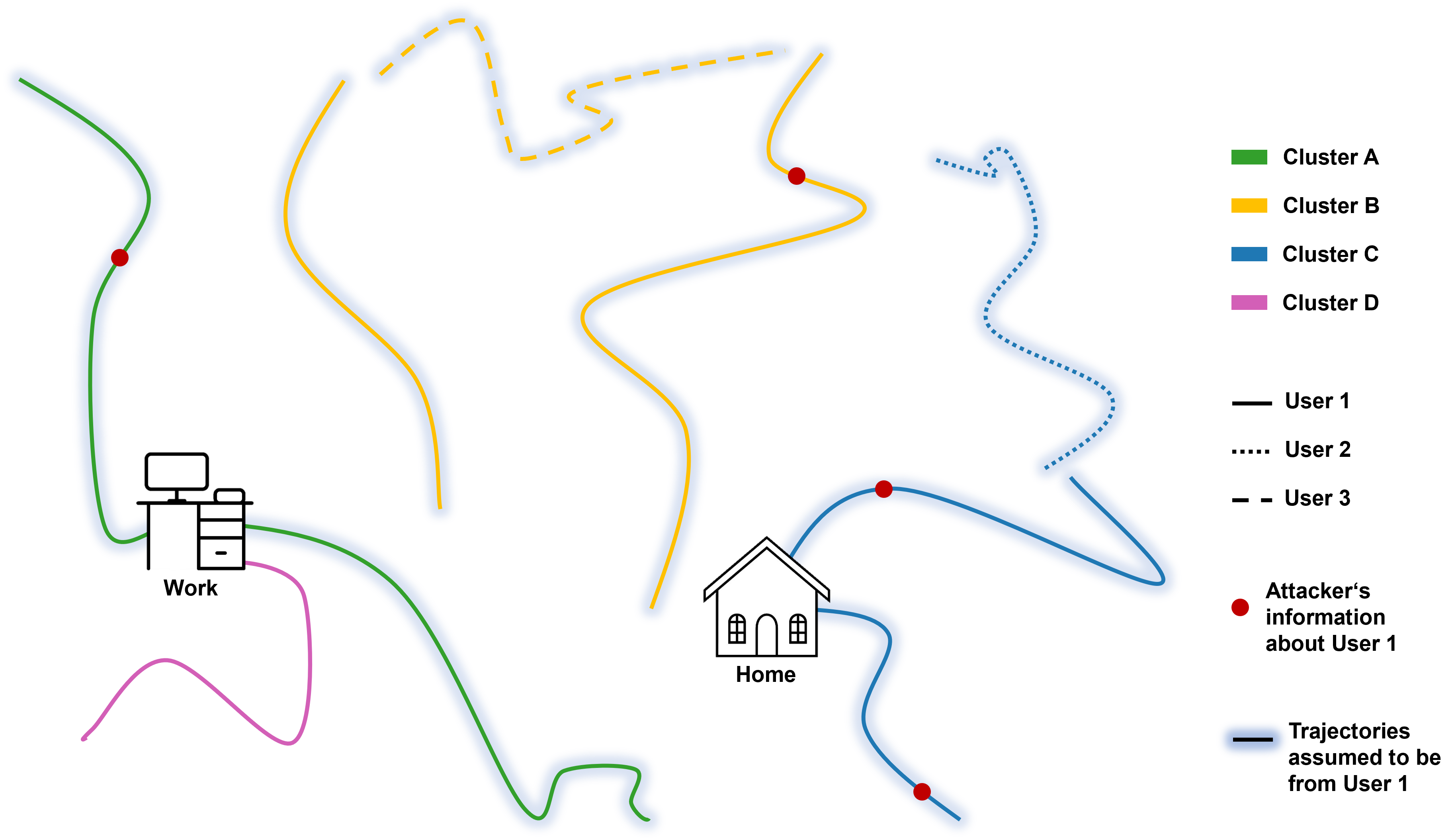}
    \caption{Attack evaluation procedure with information of four random points known to the attacker. In this case, the four known points belong to three different clusters. The attacker thus assumes the corresponding 8 trips to belong to User 1, though only 6 are correct (true positives), resulting in a precision of $\frac{6}{8}=0.75$. One trajectory of User 1 is missed (false negatives), thus resulting in a recall of $\frac{6}{7}=0.86$.}
    \label{Figure:clusteringresult}
\end{figure}

Furthermore, we evaluate the risk of re-identification for individuals by considering a scenario in which an attacker knows a certain number of locations a target person has visited. Formally, we consider the attacker's background knowledge as a set $L_p$ of $p$ randomly chosen spatio-temporal points part of the trips produced by $u \in U$. As illustrated in Figure \ref{Figure:clusteringresult}, this set of points can be used by the attacker to identify trips containing these points and, hence, all trips that are assigned to the reconstructed user IDs linked to these trips. Consequently, by combining these clusters, an attacker can create the final set of trips that they would presume to originate from the target person. 
Drawing on the approach of \citet{demontjoyeUniqueCrowdPrivacy2013b}, we set $p=4$ for our evaluation.
We consider this a low and thus reasonable amount of background knowledge, however, four spatio-temporal points have shown to uniquely identify most individuals \citep{demontjoyeUniqueCrowdPrivacy2013b}, thus posing a substantial privacy threat for mobility data (linked by user IDs).

To quantify the `correctness' of the resulting trips, we calculate the precision, i.e., the fraction of true positives to all classified positives, and recall, i.e., the fraction of true positives to all actual positives.
An example of how these metrics are calculated for one draw of $p=4$ random points from a user is shown in Figure \ref{Figure:clusteringresult}. In addition to precision and recall, we also report the F-Score in order to represent both metrics symmetrically in one score as is common practice for information retrieval problems \citep{kentMachineLiteratureSearching1955, manningIntroductionInformationRetrieval2008, najjarTrajectoryUserLinkingEasier2022a, maypetryMARCRobustMethod2020a}.

This re-identification attack evaluation is conducted for each user across 100 random samples of $L_p$ and the mean precision, recall, and F-Scores along with their 95\% confidence intervals are reported.
To rule out the possibility that the complete set of trips of a user is already included in the attacker's background knowledge, we only include users with at least $p + 1$ trips to the evaluation. Note that these users' trips remain part of the dataset when running the attack for other users, thus not increasing their re-identification success by reducing the amount of data. 

We want to point out, that we do not assume that an attacker knows the total number of users part of the dataset. Therefore, applying and evaluating the proposed attack on a transformed dataset as outlined in Chapter \ref{chapter:datasets} does not compromise the validity of such an attack since targeting a smaller dataset only limits the amount of information an attacker can gain about the individuals in the data.

\subsection{Baseline Approach}

The outlined attack is compared to the E2DTC deep clustering framework \citet{fangE2DTCEndEnd2021a}. This framework has shown state-of-the-art performance in clustering across three datasets including GeoLife. Although trajectory clustering is primarily aimed at finding spatio-temporally similar trajectories instead of underlying user behavior patterns, we chose this approach as a baseline for two reasons. First, clustering trajectories taking into account spatio-temporal similarity is likely to be correlated with finding users because urban residents' daily movements are based on routines and follow recurring routes \citep{eagleEigenbehaviorsIdentifyingStructure2009, choFriendshipMobilityUser2011b}. Second, the underlying deep representation learning method of  \citet{fangE2DTCEndEnd2021a} closely resembles the skip-gram-based idea first introduced for training word representations \citep{mikolovEfficientEstimationWord2013}, which also forms the basis of TULER and TULVAE -- two influential methods for solving semi-supervised trajectory-user linking \citep{gaoIdentifyingHumanMobility2017,zhouTrajectoryUserLinkingVariational2018}. 

We employ similar data preprocessing as outlined in Section \ref{chapter:datasets} and otherwise use the same parameters as the authors for the GeoLife dataset with a train-test split of 85\% and 15\% during pretraining of the trajectory representations. The steadily decreasing loss curves in Appendix \ref{appendix:e2dtc_curves}  show that the model is successful in learning trajectory representations for both datasets, but fails to improve on the initial clustering performance during training. This illustrates, as expected, the limited suitability of this method for clustering users only taking trajectory similarity into account. Note that we trained the baseline clustering framework by providing the correct number of target clusters $k$, which is additional side information that is not needed for our proposed attack.

\subsection{Evaluation on Raw Data}

\subsubsection{Overall performance comparison}

We run the baseline approach as well as the outlined attack on both datasets and compare their success in the following section. Overall, we observe that our attack clearly outperforms the baseline on almost all metrics and for both datasets. The results show how this attack is able to cluster trips produced by the same individual. From Table \ref{tab:overallresults}, it becomes evident that while E2DTC performs similarly on freemove and GeoLife, the method outlined in this study captures the underlying user clusters better for freemove. From the high homogeneity scores of 0.93 and 0.85, respectively, we can infer that our attack tends to be successful at avoiding the error of assigning trips of different users to the same ID. However, the resulting clusters do often not capture the entire set of trips of the users in the data as the relatively lower completeness scores indicate.

\begin{figure}[h]
    \centering
    \includegraphics[width=\linewidth]{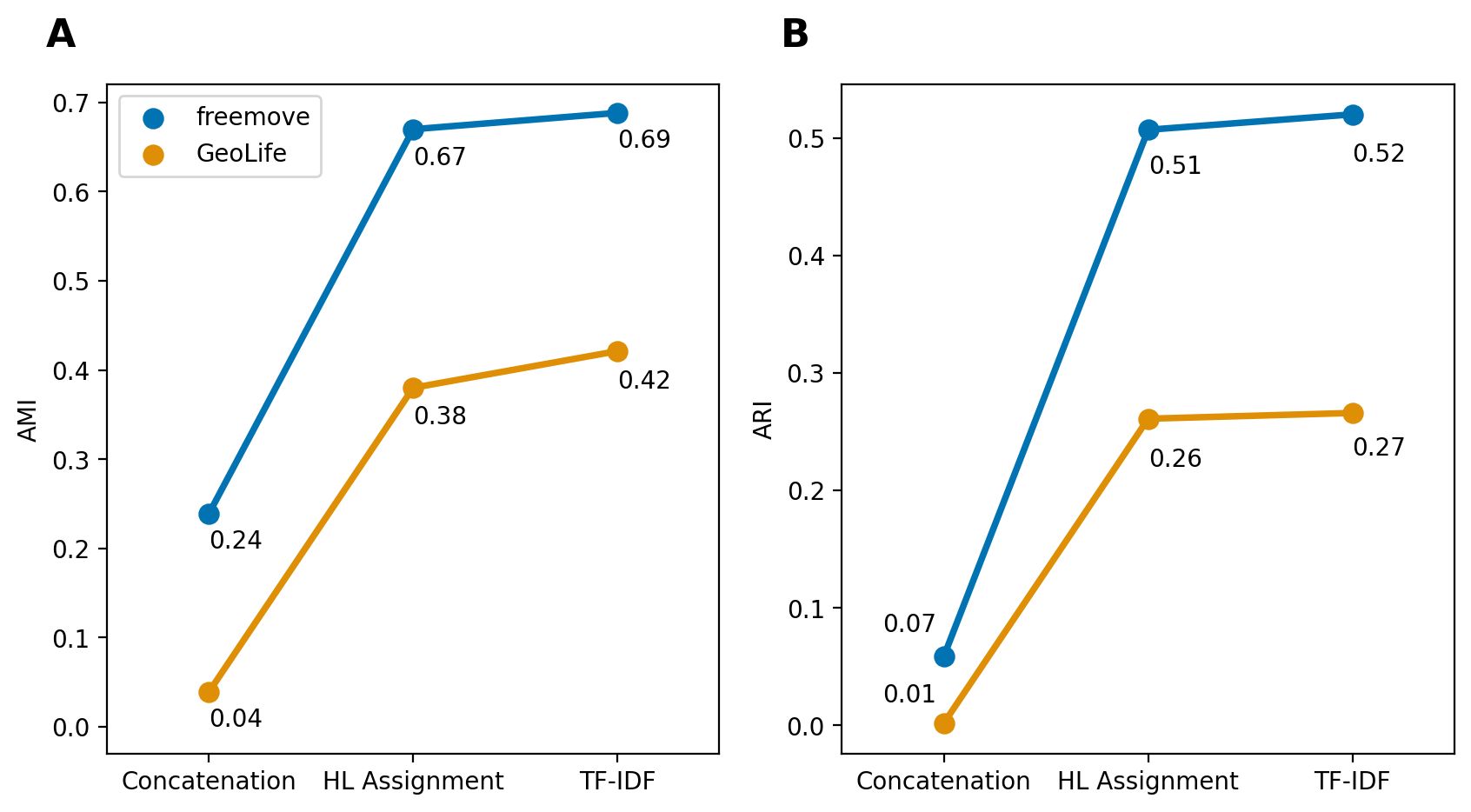}
    \caption{Incremental performance across individual heuristics of attack with respect to (A) AMI and (B) ARI.}
    \label{Figure:learningcurvesattack}
\end{figure}

\
Further, Figure \ref{Figure:learningcurvesattack} depicts the performance increments in terms of the ARI and AMI scores across the three main steps of the attack. We observe that there is a monotonous increase in overall clustering performance for both datasets, thus each step contributes to a better performance. Nevertheless, the greatest improvement in clustering performance appears to come from the heuristic to locate and assign trips to potential HLs. 
The similar pattern of both datasets indicates that the attack and its underlying assumptions are valid across different countries and data collection periods, for the considered type of mobility datasets.

\subsubsection{Characterization of re-identification risk}

\begin{figure}[h]
    \centering
    \includegraphics[width=\linewidth]{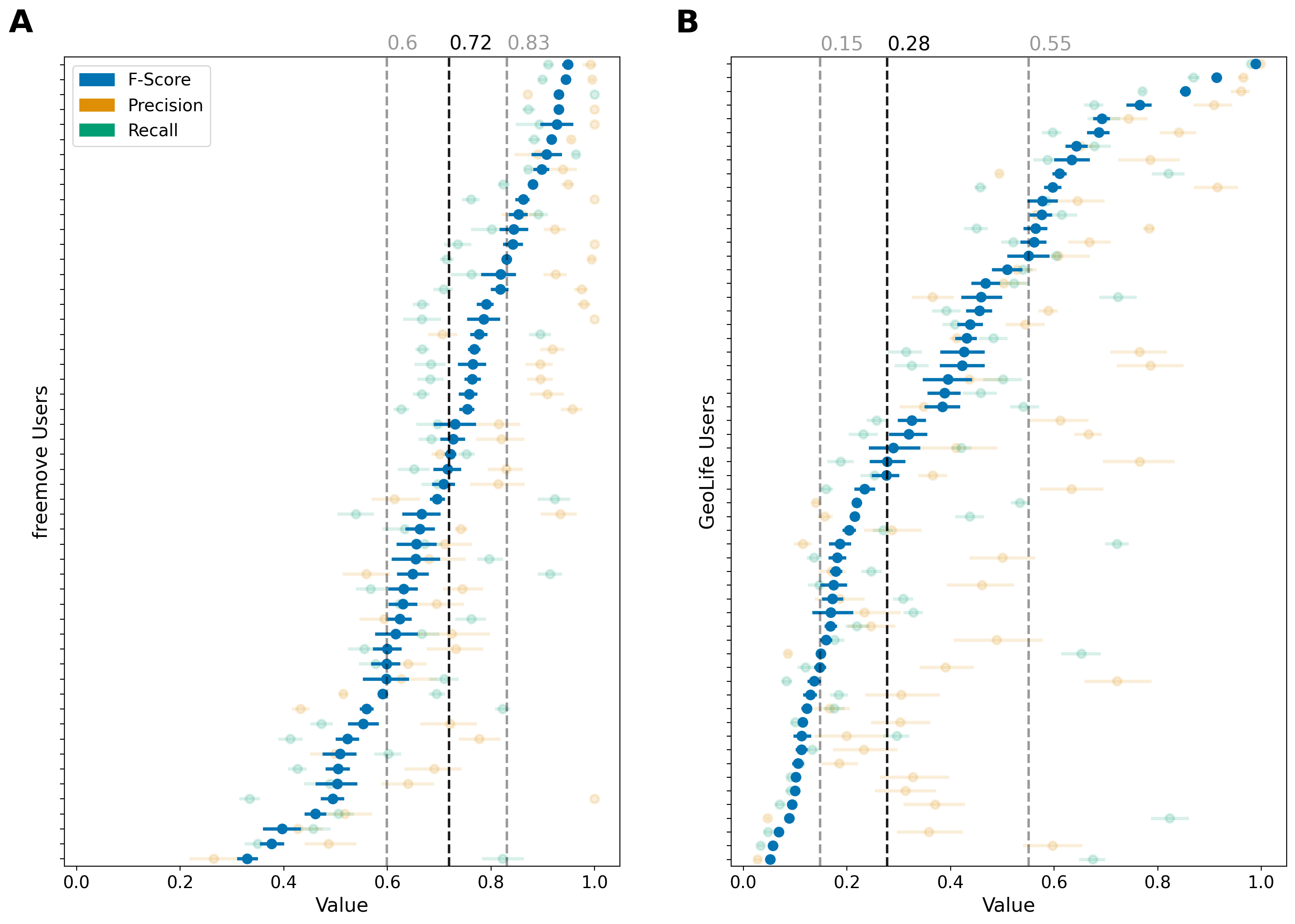}
    \caption{Mean F-score, precision, and recall across 100 random samples of $p = 4$ points for every user in (A) freemove and (B) GeoLife along with their 95\% confidence interval. Users are sorted by ascending F-score. Vertical lines indicate the median and lower and upper F-score quartiles.}
    \label{Figure:userresults}
\end{figure}

Next, we estimate the re-identification risk for individuals using $p=4$ randomly chosen spatio-temporal points from users' trips. Across 100 samples per user, we observe a median F-score of 0.72 for freemove and 0.28 for GeoLife (see Figure \ref{Figure:userresults}). It also becomes clear that analogous to homogeneity and completeness in the overall performance, precision tends to be higher than recall on both datasets. In spite of the lower median re-identification risk for users in GeoLife, Figure \ref{Figure:userresults} shows that there are users for which given $L_{p=4}$ precision and recall are very high. For the 25\% of users with the highest F-score -- the most vulnerable to our attack -- we identify the correct set of trips of individuals in freemove with an average precision of 0.96 and recall of 0.86. Calculating the same values for users in GeoLife, results in 0.77 and 0.66. In other words, this means that on average for a user in the freemove (GeoLife) dataset with an F-score in the upper 25\%, 86\% (66\%) of their trips can be correctly re-identified as part of a set with 4\% (23\%) false positives given only four random points. Considering these users' mean number of trips in the data, this would result in the re-identification of about 18 out of 21 and 75 out of 113 trips as part of a set of 19 and 97 trips for an individual in freemove and GeoLife, respectively. Overall, the results suggest that users in GPS trajectory datasets where no user ID is available, are subject to a significant risk of re-identification with some individuals being particularly vulnerable.

\subsubsection{Investigation of characteristics impacting user vulnerability}

\begin{figure*}[h]
    \centering
    \includegraphics[width=0.8\textwidth]{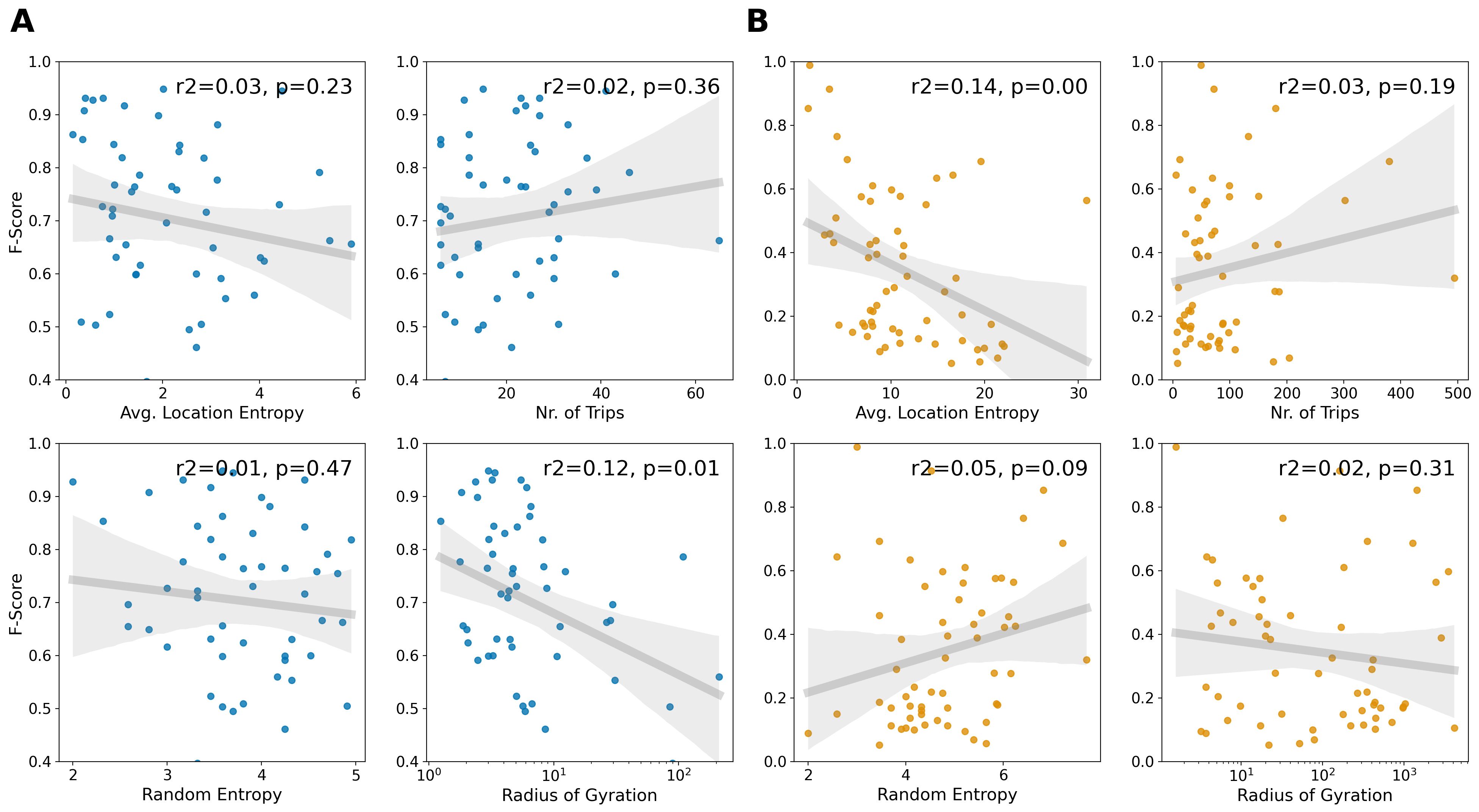}
    \caption{Mean F-scores given $p=4$ random points across mobility characteristics for users in (A) freemove and (B) GeoLife.}
    \label{Figure:usercharacteristics_nrp_4}
\end{figure*}

To investigate which factors might lead to an increased vulnerability of users to re-identification, we perform univariate linear regressions of the mean F-scores on various user-level characteristics summarizing the overall mobility behavior of individuals. Namely, these characteristics include (a) the average location entropy, which is proportional to the number of distinct users that visited a location, averaged over all locations a user has visited, (b) the random entropy, which is proportional to the number of distinct locations a user has visited, (c) the number of trips, and (d) the radius of gyration, which is the characteristic distance traveled by a user \citep{JSSv103i04}.
First, it is important to emphasize that the number of trips of a user is not significantly correlated with their risk of re-identification (see Figure \ref{Figure:usercharacteristics_nrp_4}). Further, we observe that users with lower average location entropy tend to be more vulnerable to the attack. While the effect for $L_{p=4}$ is only significant for GeoLife with $R^2=0.14$, a strong decline in p-value to 0.01 and an increase of $R^2$ to 0.16 when assuming slightly more knowledge of 10 random points suggests that this effect might hold for freemove as well (see Appendix \ref{appendix:extended_results}). Indeed, this result confirms the intuition that users traveling to locations with a lower variability of visits tend to have a higher risk of re-identification. The radius of gyration does not seem to be related to increased susceptibility. Somewhat surprisingly, the results also indicated that there is no clear relationship between random entropy and the vulnerability of a user to re-identification.

\subsection{Evaluation on Obfuscated Data}

\begin{figure}[h]
    \centering
    \includegraphics[width=\linewidth]{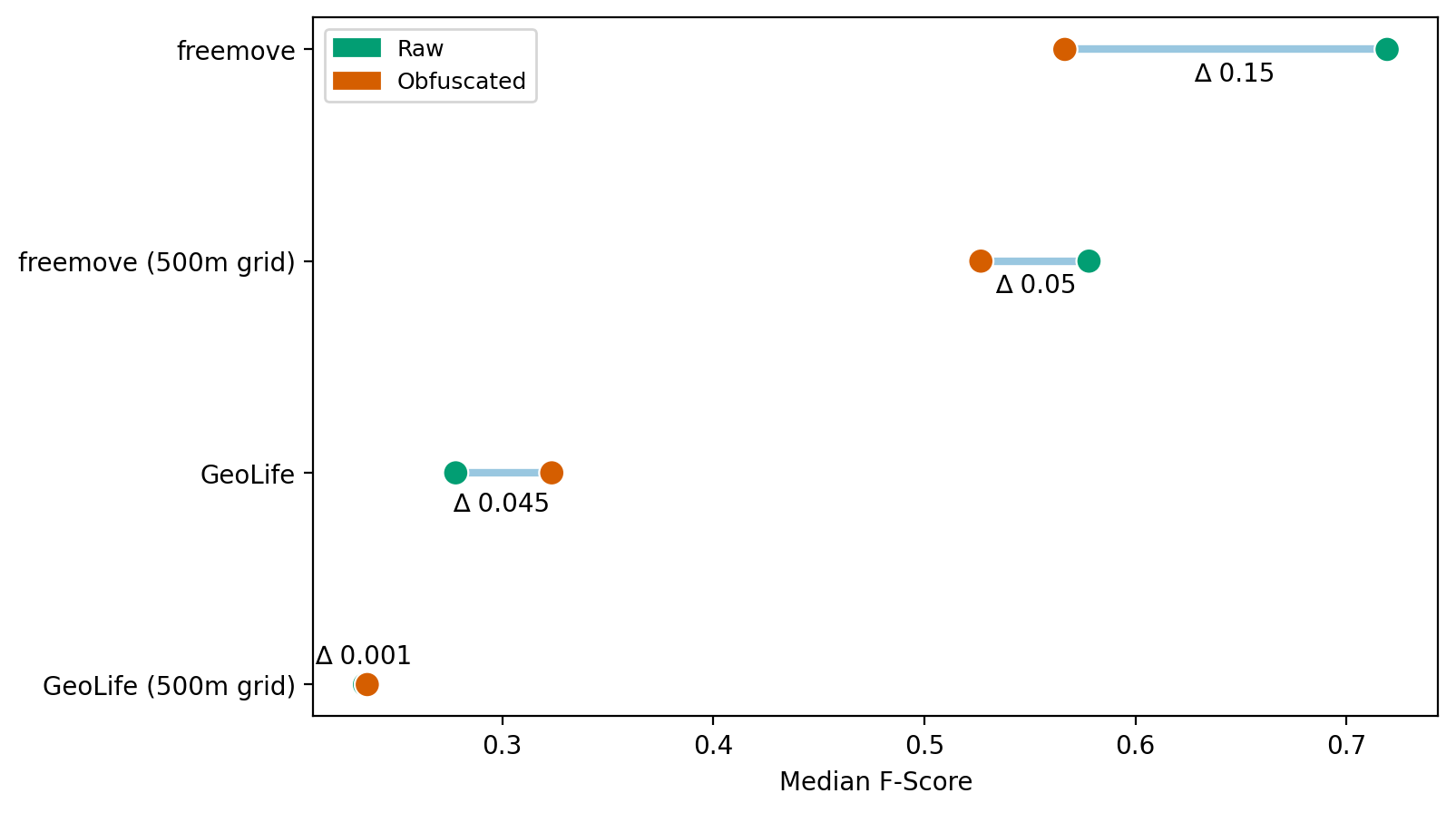}
    \caption{Median F-scores across users for freemove and GeoLife on raw and obfuscated datasets given $L_{p=4}$ with $s_{cell}=200m$ and $s_{cell}=500m$.}
    \label{Figure:obfuscateddelta}
\end{figure}

Evaluations on obfuscated data, as depicted in Figure \ref{Figure:obfuscateddelta}, show that the median F-score across users is indeed decreasing for the freemove dataset. The median F-score is 0.15 lower after truncating trips than based on the raw data. However, the same does not hold for the GeoLife dataset. Here, the difference in performance comparing raw and modified data is minimal and even amounts to a 0.045 higher median F-score on the obfuscated trips. These results strongly emphasize that the effect of truncation as a privacy-enhancing technique might increase privacy but is not necessarily robust across datasets.

%This finding is further corroborated when taking into account larger grid sizes for $s_{cell}$. 
An attacker -- if they have knowledge about the privacy mechanism as can be often assumed -- might try to counteract and deliberately choose a larger grid size to `capture' the extra noise with regard to trip lengths. Therefore, we also evaluate the attack on a grid with a cell size of 500m. The results in Figure \ref{Figure:obfuscateddelta} show first of all, that results decrease with coarser grid resolutions, thus, coarsening the grid can generally not counteract the obfuscation technique. However, we do observe for freemove %for both datasets 
that the difference in performance between raw and obfuscated data is shrinking, hence reducing the impact of the obfuscation technique. The same effect is not observed for Geolife suggesting that coarsening the grid does not have a clear effect on the attack performance.
%Hence, this evidence is further questioning the effectiveness of countermeasures such as the one applied in this study and underlines their limited robustness in real-world scenarios. 

\section{Discussion}\label{chapter:discussion}

With this work, we demonstrated that the privacy of individuals in GPS trajectory datasets is not fully protected by only removing their personal identifiers. Even assuming that an attacker has only very little outside information, there is a high risk of re-identification for a significant fraction of individuals. 

Investigating the characteristics of users, it has become clear that individual behavior patterns affect privacy risk. Our results indicate that users with a lower average location entropy across their distinct visited locations, meaning, people who frequently visit locations that are only visited by a small number of other individuals, tend to be more vulnerable to re-identification. This suggests that recording trips to personal locations, not frequently visited by the larger urban population, makes a person particularly susceptible to the attack.

While the variability in F-scores across users in both datasets is relatively high, this technically simple approach has emphasized the privacy risk inherent to GPS trajectory data. Under the assumption that technical tools will continue to evolve, one can expect more powerful attacks to be developed in the future. This attack should thus be interpreted as a starting point forming a baseline for technically more sophisticated methods, and should be used as part of penetration tests before sharing or releasing a dataset.

In addition, the evaluation of a widely-used privacy mechanism used by real-world data donation projects underlines its unreliable efficacy in protecting users' privacy. 
%The effect across both datasets is found to be sensitive and is even further mitigated when a larger grid cell size is used. 
Additionally, it should be noted that the chosen parameters are set based on findings in the extant literature and informed intuition. Further fine-tuning might improve the attack performance and reduce the impact of such a privacy mechanism. We leave the analysis of how the optimal parameter values might vary across contexts and datasets for future work.

\subsection{Generalizability of the Results}

Our findings show that the overall risk of re-identification differs between the two datasets. This raises the question of generalizability, with implications on the robustness of the attack on other datasets gathered from different contexts.

Comparing the overall characteristics between freemove and GeoLife, it becomes clear that they differ in the number of trips per user and the time period during which trips have been recorded. As for GeoLife, a longer time period will likely lead to an increase in the number of trips and unique places visited per user. Yet, we found that users' number of trips is not related to the performance of our attack while the effect of random entropy does not become entirely clear considering the scenario with the attacker knowing ten random points. This suggests that the generalizability to larger datasets depends to some extent less on the sheer size than the individual behavior characteristics of users. Consequently, we cannot conclude that users part of larger datasets will be \textit{a priori} better protected from re-identification.

The type and diversity of individuals part of a dataset are also likely to affect the results. Both, GeoLife and freemove, mostly contain individuals of one specific social group (i.e., Microsoft employees and students) and have limited representativeness in approximating a larger urban population. While this is often the case for mobility datasets \citep{luUnderstandingRepresentativenessMobile2017, wesolowskiImpactBiasesMobile2013}, it suggests that the overlap in jointly visited locations across users and, thus, the average location entropy might decrease for datasets containing a more diverse cohort of participants. This implies that users part of more diversely sampled datasets, which will continue to be simpler to collect.
With the recent developments in the regulatory landscape, data donation projects that require effective privacy solutions will be implemented at an ever-increasing rate. A report on behalf of the German government, for example, has emphasized the importance of setting up urban data platforms where administrators can collect and share mobility data relevant to local policy-makers \citep{federalinstituteforresearchonbuildingurbanaffairsandspatialdevelopmentDataStrategiesCommon2021}. Gathering these data will also become more cost-efficient and technically simpler since legislation such as the GDPR and EU's Data Governance Act have equipped users with the right to receive copies of their personal information stored by digital platforms. Therefore, projects like SimRa could soon be realized even without designing an entire app but through acquiring data recorded by platforms directly, which drastically reduces the complexity of gathering sensitive information at a larger scale \citep{boeschotenDigitalTraceData2020b, ohmeDigitalDataDonations2022a, kingEnsuringDataRichFuture2011a}. Given that donors of data collected in such a way do not need to remember, for example, to start the recording when leaving the house, it is likely that the resulting datasets are also sampled more diversely and with a higher degree of completeness. With this, as my results indicate, privacy risks of mobility data will further increase.

\subsection{Limitations}

As pointed out in the previous section, participants in both, freemove and GeoLife, are made up of a rather homogeneous group of individuals. While this does not result in reduced re-identification difficulty (see Chapter \ref{chapter:datasets}), it is still important to assess the generalizability of the results on additional mobility datasets gathered from different settings. Another consideration is the practicability of assuming a random set of points available to an attacker. As has been pointed out before, modeling a more elaborate scenario in which an attacker tries to acquire such knowledge about an individual could further improve the inferences made from studies similar to this paper \citep{pellungriniModelingAdversarialBehavior2022a, demontjoyeUniqueCrowdPrivacy2013b}. We note that the final interpretation of the privacy risks based on our findings is somewhat subjective. As pointed out by \citet{cohenPrivacyPerceptionAdolescents2014}, the individual value attributed to privacy is dependent on demographic factors and can be mitigated by a greater perceived benefit from an application. Lastly, the parameters (see Table \ref{tab:parameters}) used in this study are chosen based on common sense and extant findings about the mobility routines of urban residents. Nevertheless, predicting how an attacker would set these parameters in a real setting remains challenging and might impact this work's generalizability.

In this paper, we estimate the risk of re-identification for individuals part of GPS trip datasets with no user-trajectory link. We do so by developing a new attack methodology to reconstruct user identifiers in mobility micro-data and define the underlying trip clustering problem formally. This attack comprises a series of heuristics grounded in assumptions about the daily mobility behavior of urban residents using no outside information other than the spatio-temporal trips themselves. Evaluating our approach on two real-world datasets from Berlin and Beijing, we show that four random points are enough to re-identify a significant share of individuals' trips. Finally, we assess the efficacy of a widely-used obfuscation technique to protect users' privacy and demonstrate its limited reliability across datasets.

These results raise important concerns about the limited privacy protection achieved with simple mechanisms such as discarding the user identifier. Considering the likely development of more powerful attacks in the future, we are convinced that this work forms an important baseline for mobility privacy research. Our findings also highlight the need for a more holistic approach to privacy in location-based services that acknowledges the inherent trade-off between anonymity and utility in data-sharing.

In conclusion, the method proposed in this paper highlights the re-identification risk of GPS trip datasets, even in the absence of explicit user identifiers, and calls for further validation on diverse datasets. Moving forward, the challenges associated with unsupervised trip linking may benefit from the exploration of advanced solutions, such as deep conditional clustering methods, to expand our understanding of this critical privacy issue.

\section*{Acknowledgments}

This work is part of the freemove research project\footnote{\url{https://www.freemove.space/}}. We’d hereby thank the other project members for their valuable input, in particular, we want to thank Prof. Helena Mihaljević.

%TC:endignore

%%
%% The next two lines define the bibliography style to be used, and
%% the bibliography file.
\bibliographystyle{ACM-Reference-Format}
\bibliography{references}

%%
%% If your work has an appendix, this is the place to put it.

\newpage
\section*{Appendix}

%TC:ignore

\subsection{SimRa User Interface}

\begin{figure}[b]
    \centering
    \includegraphics[width=5cm]{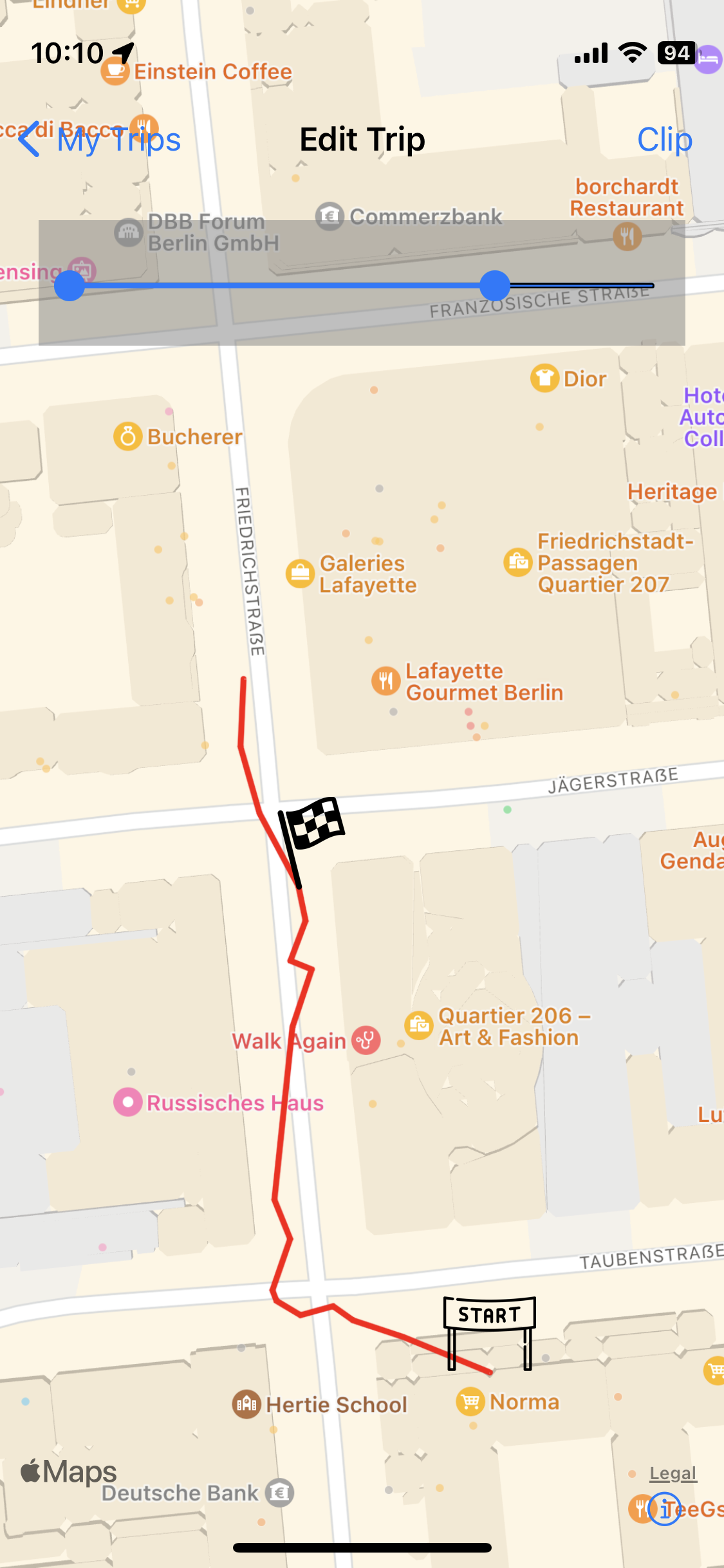}
    \caption{Screenshot of the SimRa app where users have the choice to truncate their trajectories by a variable margin set with a slider control around starting and end points.}
    \label{Figure:simratrip}
\end{figure}

\begin{figure}[b]
    \centering
    \includegraphics[width=5cm]{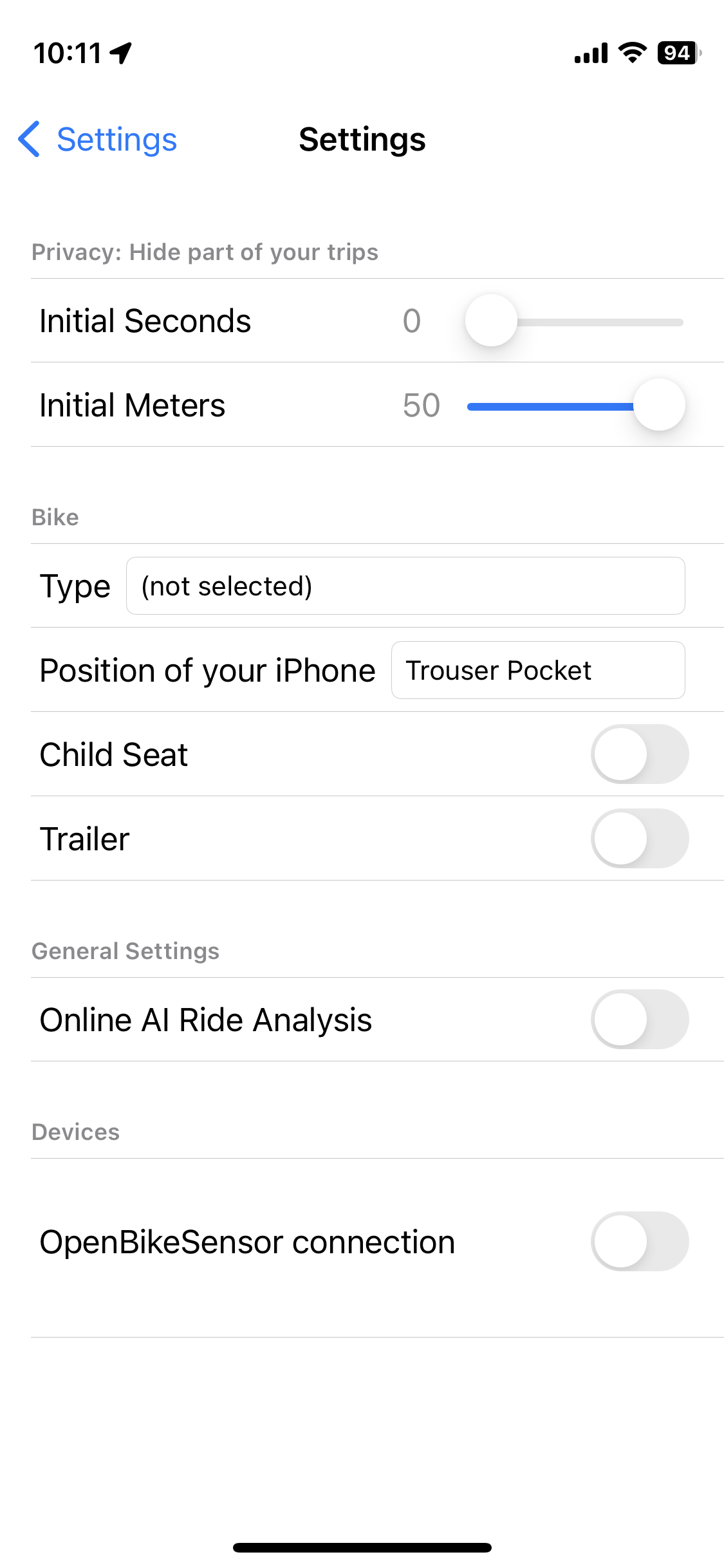}
    \caption{Screenshot of the SimRa app's preferences showing the setting for truncating an initial segment of flexible length up to 50m for every trajectory by default.}
    \label{Figure:simrasettings}
\end{figure}

Figure \ref{Figure:simratrip} and \ref{Figure:simrasettings} show screenshots of the SimRa User Interface.

\subsection{Baseline Loss Curves}
\label{appendix:e2dtc_curves}

\begin{figure}[b]
    \centering
    \includegraphics[width=\linewidth]{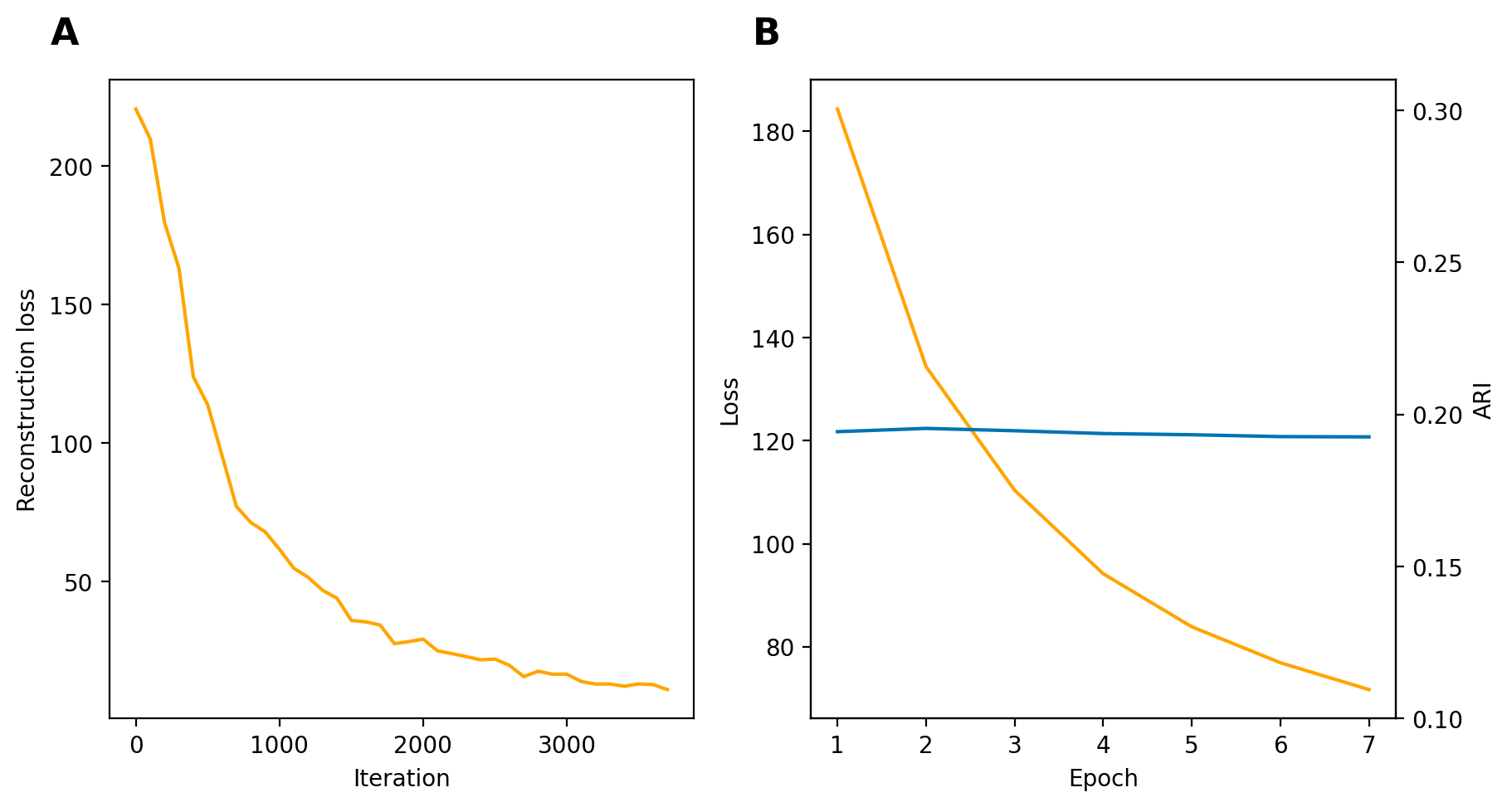}
    \caption{Loss curves and ARI (blue) for the E2DTC deep clustering framework and the freemove dataset. Plots show loss curves (orange) during (A) trajectory representation learning and (B) cluster refinement step. The stagnating clustering performance indicates that the framework fails to capture the underlying user profiles from clustering spatio-temporal similar trajectories.}
    \label{Figure:e2dtc_freemove_curves}
\end{figure}

\begin{figure}[b]
    \centering
    \includegraphics[width=\linewidth]{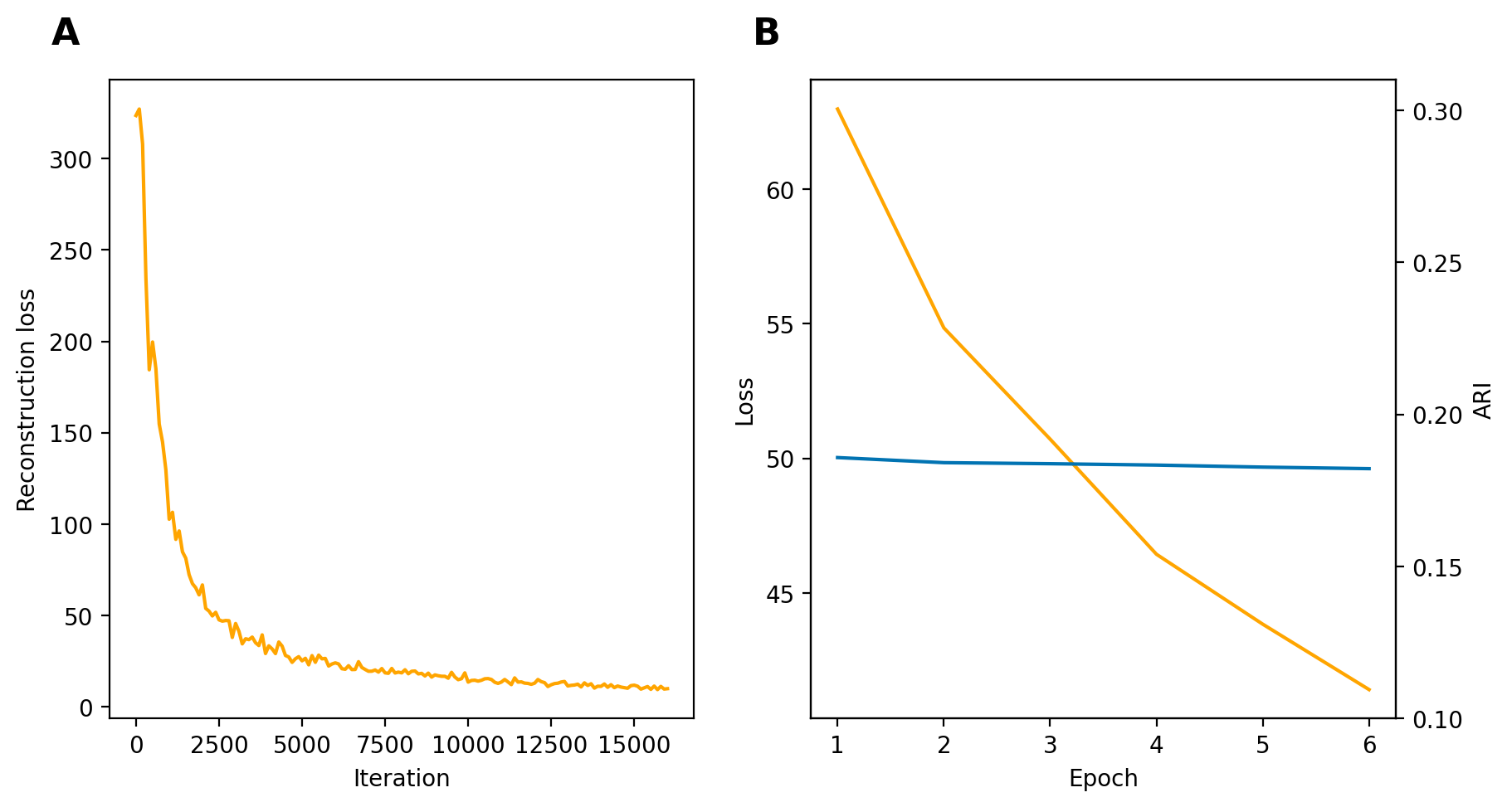}
    \caption{Loss curves and ARI (blue) for the E2DTC deep clustering framework and the GeoLife dataset. Plots show loss curves (orange) during (A) trajectory representation learning and (B) cluster refinement step. The stagnating clustering performance indicates that the framework fails to capture the underlying user profiles from clustering spatio-temporal similar trajectories.}
    \label{Figure:e2dtc_geolife_curves}
\end{figure}

 Figure \ref{Figure:e2dtc_freemove_curves}  shows the baseline loss curves for the freemove dataset and Figure \ref{Figure:e2dtc_geolife_curves} for GeoLife, respectively.

\subsection{Extended Evaluation Results}
\label{appendix:extended_results}

\begin{figure}[b]
    \centering
    \includegraphics[width=\linewidth]{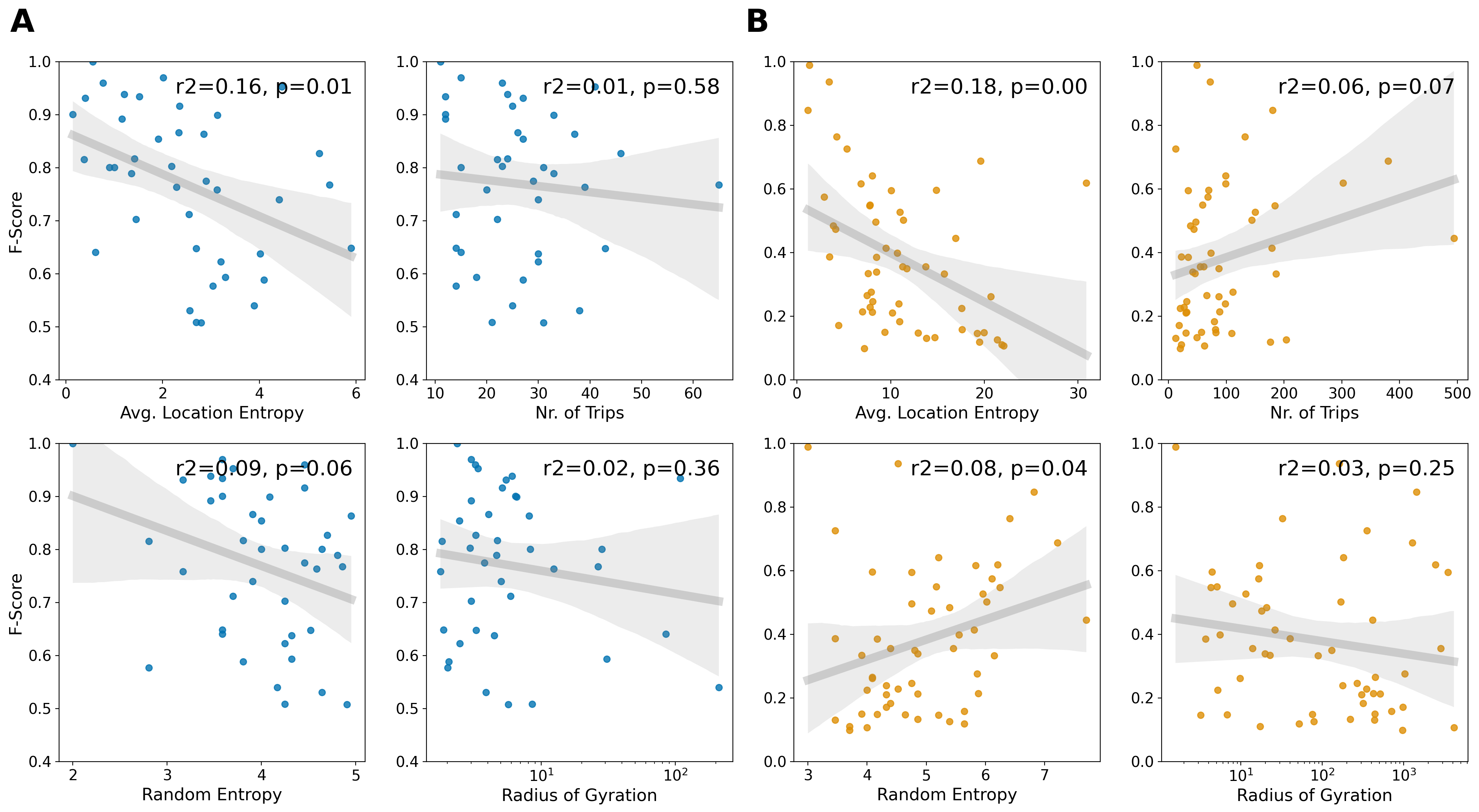}
    \caption{Mean F-scores given random points $L_{p=10}$ across mobility characteristics for users in (A) freemove and (B) GeoLife.}
    \label{Figure:usercharacteristics_nrp_10}
\end{figure}

\begin{figure}[b]
    \centering
    \includegraphics[width=\linewidth]{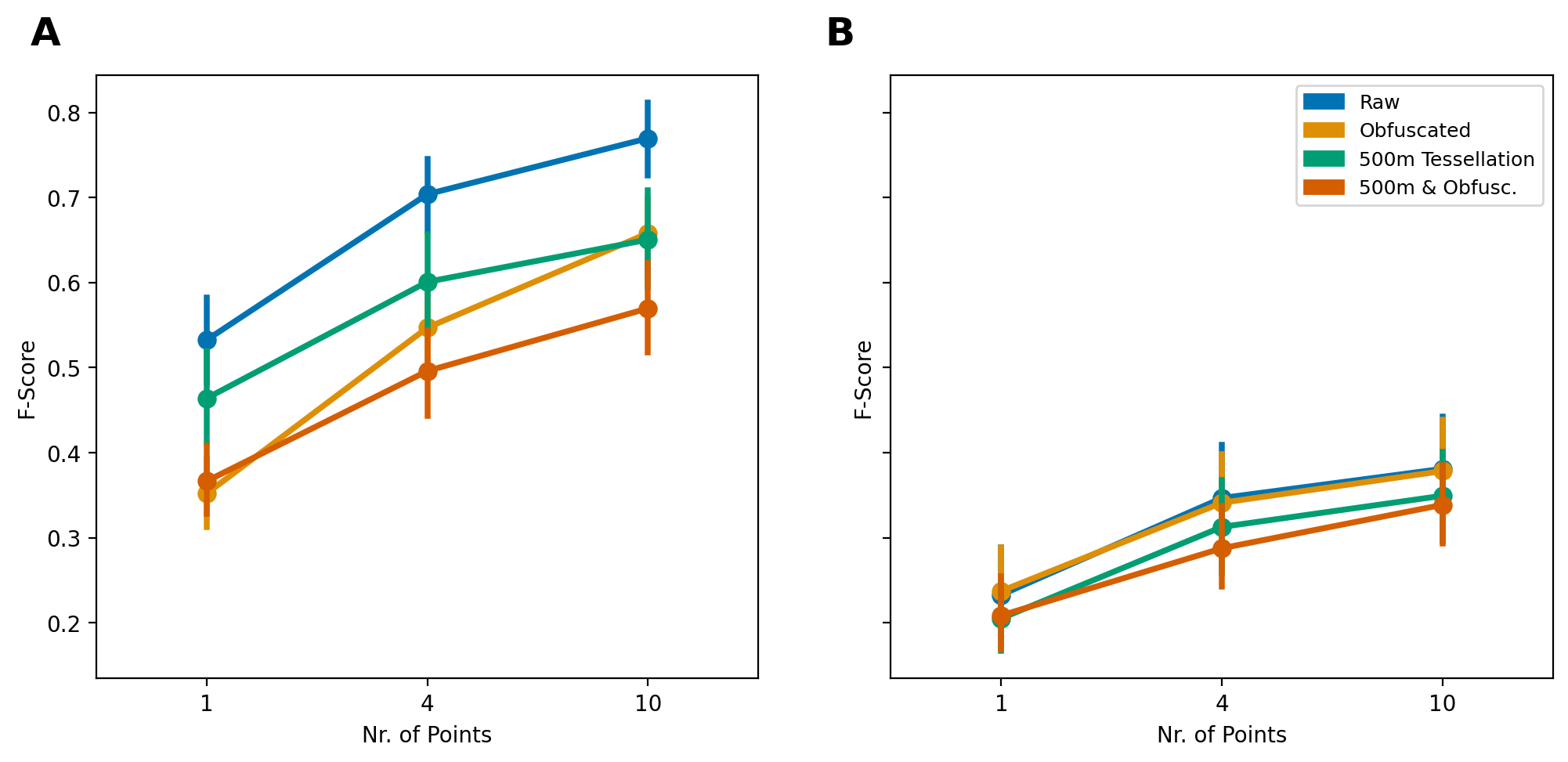}
    \caption{Performance across data types and grid cell sizes for various numbers of random spatio-temporal points assumed as knowledge of an attacker. (A) freemove and (B) GeoLife.}
    \label{Figure:obfuscated}
\end{figure}

Figure\ref{Figure:usercharacteristics_nrp_10}  shows evaluation results for $L_{p=10}$ and Figure \ref{Figure:obfuscated} performance results for different grid cell sizes with and without additionally applied obfuscation technique.

\newpage

\section{Tables}

\begin{table}[ht]
\begin{center}
\small
\begin{tabular}{p{0.4cm}p{3cm}p{0.6cm}rrrr}
  \toprule
   &  & \textbf{min} & \textbf{1. quart} & \textbf{med} & \textbf{3. quart} & \textbf{max}\\
  \midrule
  \textbf{fm} & trips per user & 1.0 & 5.5 & 13.0 & 26.3 & 65.0 \\
  \textit{} & length (km) & 0.2 & 1.5 & 4.3 & 9.4 & 24.0 \\
  \textit{} & duration (min) & 1.1 & 12.8 & 25.7 & 42.5 & 259.1 \\
  \textit{} & cells per user & 1.0 & 7.0 & 13.0 & 23.3 & 38.0 \\
  \textbf{GL} &trips per user & 1.0 & 11.3 & 46.5 & 87.3 & 494.0 \\
  \textit{} & length (km) & 0.2 & 2.8 & 8.3 & 18.3 & 81.9 \\
  \textit{} & duration (min) & 1.4 & 17.2 & 42.1 & 158.7 & 2085.3 \\
  \textit{} & cells per user & 2.0 & 14.8 & 34.5 & 68.3 & 335.0 \\
  \bottomrule
\end{tabular}
\caption{Five-number summary of trips per user, trip length, trip duration, and cells per user for both datasets freemove (fm) and GeoLife (GL).} \label{tab:datasummary}
\end{center}
\end{table}

\begin{table}[ht]
\begin{center}
\begin{tabular}{lr}
  \toprule
  \textbf{parameter} & \textbf{value} \\
  \midrule
  $s_{cell}$ & 200m \\
  $h_{concat}$ & 8 \\
  $[h_{concat,before}$, $h_{concat,after}]$ & [-4, 4] \\
  $t_{morning}$ & [6:00 AM, 10:00 AM] \\
  $t_{evening}$ & [6:00 PM, 12:00 AM] \\
  $[h_{morning, before}, h_{morning, after}]$ & [-2, 2] \\
  $[h_{evening, before}, h_{evening, after}]$ & [0, 4]\\
  $LCSS_{\varepsilon}$ & 200m\\
  $s_{cell, tfidf}$ & 500m\\
  $n_{matches}$ & 5 (freemove); 100 (GeoLife) \\
  $q_{match}$ & 0.75 \\
  \bottomrule
\end{tabular}
\caption{Overview of parameter values used throughout the attack.} \label{tab:parameters}
\end{center}
 \end{table}
 
\begin{table}[h]
\begin{center}
\begin{tabular}{llrr}
  \toprule
    & \textbf{metric} & \textbf{our approach} & \textbf{E2DTC} \\
  \midrule
  \textbf{freemove} & ARI & \textbf{0.52} & 0.19 \\
  \textbf{} & AMI & \textbf{0.69} & 0.53 \\
  \textbf{} & Homogeneity & \textbf{0.93} & 0.57 \\
  \textbf{} & Completeness & \textbf{0.74} & 0.53 \\
  \textbf{GeoLife} & ARI & \textbf{0.27} & 0.19 \\
  \textbf{} & AMI & 0.42 & \textbf{0.52} \\
  \textbf{} & Homogeneity & \textbf{0.85} & 0.56 \\
  \textbf{} & Completeness & \textbf{0.51} & 0.49 \\
  \bottomrule
\end{tabular}
\caption{Overall clustering evaluation results of this approach compared with the deep clustering baseline.} \label{tab:overallresults}
\end{center}
\end{table}

\clearpage

\section{Figures}

\begin{figure}[b]
    \centering
    \includegraphics[width=\linewidth]{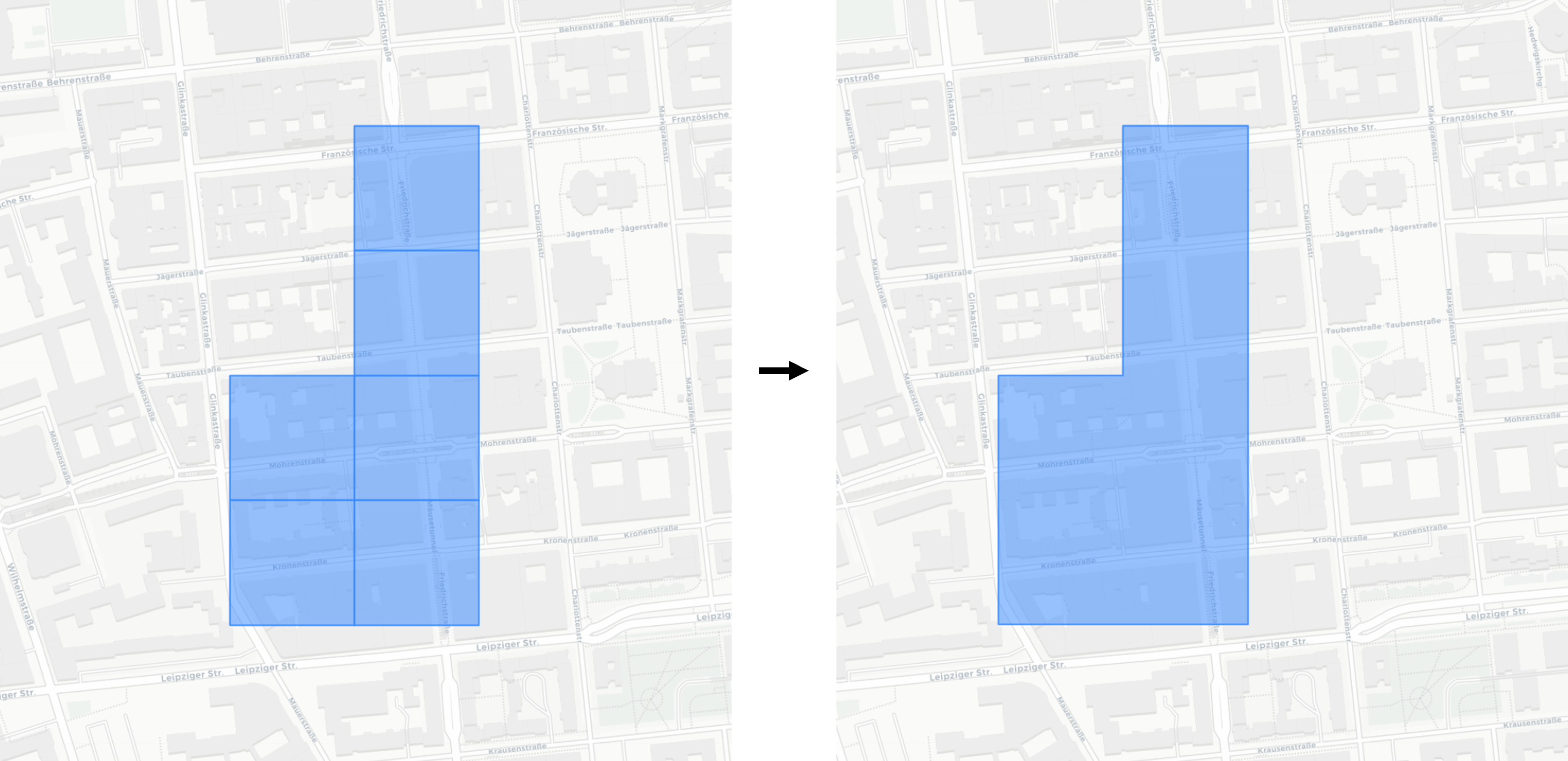}
    \caption{Example illustration of adjacent HL grid cells that would be dissolved into one combined polygon during the HL assignment step.}
    \label{Figure:dissolvingexample}
\end{figure}

\label{appendix:simra}

%TC:ignore

%TC:endignore

%TC:ignore

%TC:endignore

\end{document}